\documentclass[oneside,twocolumn,aps,pra,preprintnumbers,bibnotes10pt,superscriptaddress,amsmath,amssymb]{revtex4-2}
\usepackage{graphicx,subfigure,wrapfig}
\usepackage{float}
\usepackage{color}
\usepackage{epstopdf}
\usepackage{amsmath}
\usepackage{braket}
\usepackage{amsthm}
\usepackage{amssymb}
\usepackage{amsfonts}
\usepackage{graphicx,subfigure}
\usepackage{txfonts}
\usepackage{ulem}
\usepackage{mathtools}
\usepackage{bm}
\usepackage{hyperref}
\usepackage{natbib}
\usepackage{comment}
\usepackage[dvipsnames]{xcolor}

\newcommand{\be}{\begin{equation}}
\newcommand{\ee}{\end{equation}}
\newcommand{\beq}{\begin{eqnarray}}
\newcommand{\eeq}{\end{eqnarray}}
\newcommand{\vect}[1]{\bm{#1}}

\begin{document}
%
%\title{Increased stability of a persistent current in a toroidal superfluid with controlled impurities}
%\title{Increased stability of persistent current in a toroidal superfluid with controlled impurities}
%

\title{Tuning the Critical Current in Toroidal Superfluids via Controllable Impurities}

\author{K. Xhani}

\affiliation{Istituto Nazionale di Ottica del Consiglio Nazionale delle Ricerche (CNR-INO), Largo Enrico Fermi 6, 50125 Firenze, Italy}
\affiliation{European Laboratory for Nonlinear Spectroscopy (LENS), Via N. Carrara 1, 50019 Sesto Fiorentino, Italy}
\affiliation{Department of Applied Science and Technology (DISAT), Politecnico di Torino, Torino, Italy}

\author{G. Del Pace}

\affiliation{Istituto Nazionale di Ottica del Consiglio Nazionale delle Ricerche (CNR-INO), Largo Enrico Fermi 6, 50125 Firenze, Italy}
\affiliation{European Laboratory for Nonlinear Spectroscopy (LENS), Via N. Carrara 1, 50019 Sesto Fiorentino, Italy}
\affiliation{University of Florence, Physics Department, Via Sansone 1, 50019 Sesto Fiorentino, Italy}
\affiliation{INFN, Sezione di Firenze, 50019 Sesto Fiorentino, Italy}

\author{N. Grani}
\affiliation{Istituto Nazionale di Ottica del Consiglio Nazionale delle Ricerche (CNR-INO), Largo Enrico Fermi 6, 50125 Firenze, Italy}
\affiliation{European Laboratory for Nonlinear Spectroscopy (LENS), Via N. Carrara 1, 50019 Sesto Fiorentino, Italy}
\affiliation{INFN, Sezione di Firenze, 50019 Sesto Fiorentino, Italy}

\author{D. Hernández-Rajkov}
\affiliation{Istituto Nazionale di Ottica del Consiglio Nazionale delle Ricerche (CNR-INO), Largo Enrico Fermi 6, 50125 Firenze, Italy}
\affiliation{European Laboratory for Nonlinear Spectroscopy (LENS), Via N. Carrara 1, 50019 Sesto Fiorentino, Italy}
\affiliation{INFN, Sezione di Firenze, 50019 Sesto Fiorentino, Italy}

\author{B. Donelli}
\affiliation{Istituto Nazionale di Ottica del Consiglio Nazionale delle Ricerche (CNR-INO), Largo Enrico Fermi 6, 50125 Firenze, Italy}
\affiliation{European Laboratory for Nonlinear Spectroscopy (LENS), Via N. Carrara 1, 50019 Sesto Fiorentino, Italy}

\author{G. Roati}
\affiliation{Istituto Nazionale di Ottica del Consiglio Nazionale delle Ricerche (CNR-INO), Largo Enrico Fermi 6, 50125 Firenze, Italy}
\affiliation{European Laboratory for Nonlinear Spectroscopy (LENS), Via N. Carrara 1, 50019 Sesto Fiorentino, Italy}
\affiliation{INFN, Sezione di Firenze, 50019 Sesto Fiorentino, Italy}

\author{L. Pezzè}
\affiliation{Istituto Nazionale di Ottica del Consiglio Nazionale delle Ricerche (CNR-INO), Largo Enrico Fermi 6, 50125 Firenze, Italy}
\affiliation{European Laboratory for Nonlinear Spectroscopy (LENS), Via N. Carrara 1, 50019 Sesto Fiorentino, Italy}
\begin{abstract}
We combine numerical and experimental approaches to study how impurities affect the maximum superflow in an annular Bose–Einstein condensate. By tuning the impurity density, we achieve precise control over the stability of persistent currents which increases with the impurity number. In the unstable regime, the complex vortex motion within the impurity landscape, characterized by pinning and unpinning events, governs the timescale of the current decay and its final value.
Our work establishes atomic superfluids as a pristine platform for exploring universal mechanisms of superflow stabilization and decay, paving the way for atomtronic quantum technologies.
\end{abstract}

\maketitle

{\it Introduction.---}
Impurities in superfluids and superconductors play a multifaceted role~\cite{Anderson1959, Abrikosov1957}.
On the one hand, they facilitate vortex nucleation and the associated dissipation, converting nominally lossless flow into resistive dynamics~\cite{Tinkham1996, Blatter1994}.
On the other hand, impurities enhance stability by acting as pinning centers that immobilize vortices and suppress energy losses~\cite{Larkin1979}. 
Vortex pinning is essential for type-II superconductors to sustain dissipationless currents~\cite{Campbell1972, Brandt1995}, while in neutron stars the sudden unpinning of a large number of vortices from the nuclear lattice is thought to trigger the abrupt changes in rotation rate known as pulsar glitches~\cite{Anderson1975, Haskell2015}.
This subtle interplay underlies a broad range of transport phenomena, such as flux flow resistivity, critical currents, phase coherence, and the onset of turbulence~\cite{KopninRPP2002, Vinen2002, Barenghi2014}. 
Engineering controllable impurities provides a powerful tool to tailor superconducting and superfluid properties, from boosting vortex pinning and critical currents to selecting otherwise unstable phases~\cite{Maiorov2009, Dmitriev2015, Wordenweber2017, Neverov2022, Gastiasoro2018, Leroux2019, Nguyen2024}.
A key challenge is to identify which impurity attributes (strength, width, magnetic properties, etc) and spatial correlations control phase stability, transport, coherence, and how they affect vortex nucleation~\cite{Kwok2016}.
Addressing this problem demands 
advanced numerical simulations to accurately capture the complexity of the disorder present in real materials. 
An alternative strategy is to engineer platforms where impurities and superfluid dynamics can be precisely controlled and numerical simulations can be performed efficiently.  

Ultracold gases provide exactly such a clean platform. 
Disorder and impurities can be engineered in a controlled manner, for example via optical speckle potentials~\cite{BillyNATURE2008, JendrzejewskiNATPHYS2012, NaglerPRL2022}, incommensurate potentials~\cite{RoatiNATURE2008}, or programmable phase masks devices~\cite{WhiteNATCOMM2020}.
In ring traps, the superflow can be realized in ultraclean conditions, with the multiply connected geometry supporting metastable persistent current states characterized by quantized winding numbers~\cite{RamanathanPRL2011, MoulderPRA2012, WrightPRL2013, CaiPRL2022, PoloPR2025}. 
This tunability makes these systems versatile testbeds for exploring the interplay of impurity, interactions and topology in quantum fluids~\cite{EckelNATURE2014, EckelPRX2014, PezzeNATCOMM2024}, with direct atomtronic applications~\cite{PyuNATCOMM2020, AmicoRMP2022, GanARXIV}. 
Previous studies have demonstrated that introducing a single impurity in ring superfluids can induce vortex emission, with critical velocities determined by the impurity’s characteristics, the ring’s size, and with different behaviors for bosonic \cite{XhaniATOMS2023} or fermionic superfluids \cite{XhaniPRR2025, DelPacePRX2022}. 
In contrast, the richer case of multi-impurity systems has remained largely uncharted.

In this work, we exactly follows this bottom-up approach and address the interplay between superflow dissipation and vortex dynamics in the presence of controllable impurities.
Our platform consists of a toroidal atomic Bose–Einstein condensate (BEC) whose excitation spectrum is purely collective.  
We find that increasing the number of impurities stabilizes the superfluid current against decay, with no intrinsic limit. 
This stands in stark contrast to recent reports on atomic Fermi superfluids in the Bardeen–Cooper–Schrieffer regime, where Cooper-pair breaking limits the stability of persistent currents in the presence of impurities~\cite{btuzemen}.
In the case of identical impurities, we predict a linear increase of the critical circulation with impurity number. 
This behavior depends on the impurity landscape.
Below a critical threshold, the current remains stable in time, whereas above it the current decays via the symmetric emission of one vortex from each impurity.
Notably, both the decay rate and the residual current are set by vortex–impurity scattering dynamics.
In the experiment, we recover the bipartite stability diagram. 
However, the current's stability is affected by unavoidable deviations from a perfectly symmetric impurity configuration, in agreement with numerical results that include shifts in impurities position.
 
%%%%%%%%%%%%%%%%%%%%%%%%%%%
%% FIGURE 1
%%%%%%%%%%%%%%%%%%%%%%%%%%%
\begin{figure*}[t!]
\includegraphics[width=0.9\textwidth]{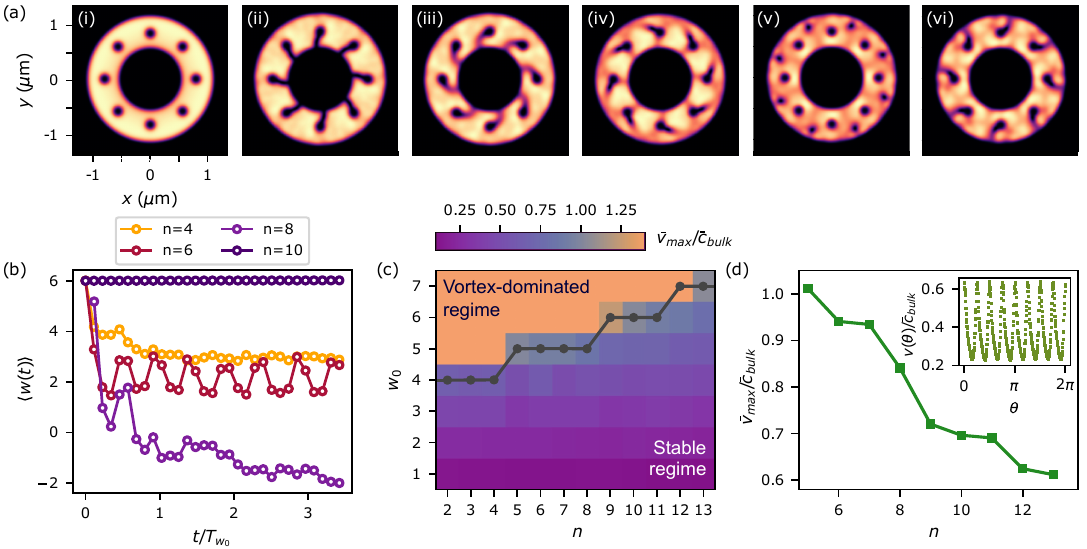}
\caption{{\bf Phase diagram of a ring superfluid with symmetric impurities}. 
(a) Snapshots of the density in the $x$-$y$ plane (integrated along the $z$ axis) at different times for the unstable dynamics with $n=8$ impurities and initial winding $w_0=6>w_c$: $t= 0 \, $ms (i), $5.7\,$ms (ii), $7.0\,$ms (iii), $8.9\,$ms (iv), $15.3\,$ms (v) and $20.3\,$ms (vi).
(b) Average winding number $\langle w(t) \rangle$ as a function of time, for different values of $n$ and fixed $w_0=6$.
The initial temporal evolution (for  $t \lesssim  0.5\, T_{w_0}$) for $n=8$ corresponds to the snapshot shown in panel (a).
Times are rescaled in rotation units~\cite{notaunits}.
(c) Stability phase diagram of the persistent current as a function of $n$ and $w_0$ in terms of $\bar{v}_{\max}/\bar{c}_{\rm bulk}$ (see text). 
The dark gray dots indicate the critical winding number $w_c$, the solid line is a guide to the eye.
(d) Time averaged superfluid velocity  $\bar{v}_{\max}/\bar{c}_{\rm bulk}$ as a function of  $n$ at fixed  $w_0=5$ and for $n \geq 5$, corresponding to the stable configurations for the chosen $w_0$.
The inset reports the angular velocity taken at fixed radius $R^*$ for $w_0$ = 5 and $n=12$.
}
\label{fig:Fig1}
\end{figure*}
%%%%%%%%%%%%%%%%%%%%%%%%%%%
%%%%%%%%%%%%%%%%%%%%%%%%%%%
%%%%%%%%%%%%%%%%%%%%%%%%%%%

\textit{The system.}---We consider a molecular BEC of $^6$Li atom pairs confined in a ring potential in the $x-y$ plane, with a tight harmonic confinement along the $z$ axis [see Appendix for details].  
The ring has width $\Delta R = R_{\rm out} - R_{\rm in} = 11.4$ $\mu$m with outer and inner radii $R_{\rm out} = 21 $ $\mu$m and $R_{\rm in} = 9.6$ $\mu$m, respectively. 
Up to $n \leq 16$ repulsive Gaussian obstacles are embedded within the ring, each  with height $V_0 = 3.7\, \mu$ and a $1/e^2$ width of $\sigma = 1.4\,\mu{\rm m} \simeq 2 \, \xi$. 
The current is initialized by imprinting a phase to the condensate wavefunction. 
This sets the initial winding number to $w_0$, such that the initial circulation (at time $t_0=0$) is
\be
\kappa(R,t_0)=\oint_{\Gamma(R)} v\, (\vect{r},t_0) \, d \vect{r}= \frac{h}{m} w_0,
\ee
with $\Gamma(R)$ being a circular loop with radius $R$ centered at the origin, $v(\vect{r},t) = (\hbar/m) \nabla\phi(\vect{r},t)$ 
the local superfluid velocity, $\phi$ the superfluid phase, and $m$ the molecule mass. 
The condensate dynamics is investigated by solving the three-dimensional time-dependent Gross-Pitaevskii equation (see Appendix~\ref{AppB} for details). 
The supercurrent stability is studied by extracting the temporal evolution of the radial-averaged winding: 
$\langle w(t) \rangle = (m/h)  \int_{R_{\rm in}}^{R_{\rm out}} \kappa(r,t) dr /\Delta R$

%%%%%%%%%%%%%%%%%%%%%%%%%%%
%% FIGURE 2
%%%%%%%%%%%%%%%%%%%%%%%%%%%
\begin{figure}[hbt!]
\includegraphics[width=\columnwidth]{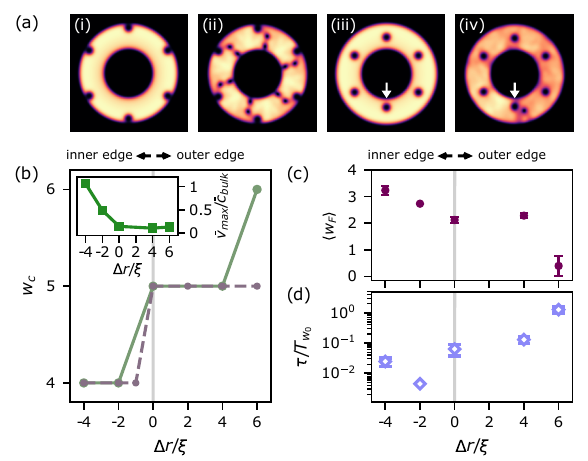}
\caption{{\bf Role of impurity distribution on the current stability}. 
(a) The two-dimensional condensate density in the $xy$-plane, initially [(i) and (iii)], and after the vortex emission [(ii) and (iv)].
The cases (i) and (ii) refer to $n=6$ impurities displaced at the ring edge, while in (iii) and (iv) only one impurity (highlighted by the white arrow) is shifted toward the inner edge, the other impurities being at mid-radius $\bar{r}$. 
(b) Critical winding number as a function of the radial position shift $\Delta r/\xi$ for $n=6$.
Green (grey) dots show the case when all (one) impurities are shifted from $\bar{r}$.
Lines are guide to the eye.
Inset: Scaled superfluid velocity $\bar{v}_{\max}/\bar{c}_{\rm bulk}$ at fixed $w_0=4$ and $n=6$, in the symmetric configuration where all impurities are displaced by $\Delta r/\xi$. 
(d-e) %Temporal evolution of the average winding number after subtracting the critical value $w_c(n)$ (dots) and its exponential fit $f(t) = a \exp(-t/\tau) + w_f$ (solid line). 
%These data are for fixed $w_0 - w_c$ and for $\Delta r/\xi =6$ (orange dots), $\Delta r/\xi =4$ (salmon dots), $\Delta r/\xi =0$ (red) and $\Delta r/\xi =-4$ (brown)  values in the periodic configuration. 
%
The final winding number $w_f$  (d) and the decay time $\tau$ in rotation units \cite{notaunits}(e) with corresponding error bars as extracted from an exponential fit of the current decay for different impurity shifts $\Delta r /\xi$ in the symmetric configuration. Here we plot the results for the first unstable winding for each impurity configuration, namely for initial winding $w_0 = w_c +1$.
}
\label{fig:Fig2}
\end{figure}
%%%%%%%%%%%%%%%%%%%%%%%%%%%
%%%%%%%%%%%%%%%%%%%%%%%%%%%
%%%%%%%%%%%%%%%%%%%%%%%%%%%

\textit{Symmetric impurities configuration.}---We begin by examining the ideal case of identical impurities symmetrically placed at the mid-radius of the torus, $\bar r=(R_{\rm in}+R_{\rm out})/2$, see Fig.~\ref{fig:Fig1}(a,i).
We find two different dynamical regimes separated by a critical circulation $w_c$: a stable regime ($w_0 \leq w_c$), where the superfluid current remains stable ($w(t)=w_0$), with no vortices emitted into the bulk; and a vortex-dominated regime ($w_0> w_c$) where the current decays to lower windings with a time scale set by the complex vortex-impurity interaction. 
An example of supercritical dynamics is shown in Fig.~\ref{fig:Fig1}(a,ii-vi). 
The onset of vortex emission corresponds to the formation of low-density channels connecting each impurity to the torus’s inner boundary [Fig.~\ref{fig:Fig1}(a,ii)], similar to the single-impurity~\cite{DelPacePRX2022,XhaniATOMS2023} or single-barrier~\cite{YakimenkoPRA2015, PiazzaPRA2009} scenario. 
Shortly after [Fig.~\ref{fig:Fig1}(a,iii)], $n$ vortices simultaneously enter the condensate through these channels, producing the initial drop in the average winding number, see Fig.~\ref{fig:Fig1}(b).
Following their entry, the vortex dynamics become strongly influenced by their interactions with impurities as well as the background superflow [Fig.~\ref{fig:Fig1}(a,iv-vi)].
The strength of vortex-impurity interactions depends on the impurity parameters, the superflow velocity, and the spatial alignment between vortex cores and impurities~\cite{StockdalePRL2021, LiuJLTP2024, DoranPRA2024}. 
For the chosen parameters, vortices tend to spiral partially around impurities before propagating outward toward the torus’s outer edge. 
With weaker impurities, vortices can pass closer to the ring's inner edge without spiraling, indicating a milder vortex-impurity interaction, as it occurs for $V_0/\mu<1$ in the single-obstacle case~\cite{XhaniATOMS2023}.

The density of impurities further shapes the vortex behavior, especially their pinning events, see Fig.~\ref{fig:Fig1}(b). 
At low impurity density ($n \leq 5$), vortices propagate without being pinned, resulting in a decaying winding number. 
For higher density of impurities (e.g.,~$n=6,7,8$) at specific initial windings, such as $w_0=6$ (i.e., larger initial superflow velocity), vortices experience temporary, weak pinning—brief absorption by impurities—before detaching.
These pinning (unpinning) events are reflected by an increase (decrease) of $\langle w \rangle $, which probes not only the superflow dissipation but also vortex motion, as vortices modify the local superfluid phase profile by causing local phase slippage. 
Remarkably, for $n=6$, weak pinning/unpinning events persists over long times and recurs periodically, generating nearly sinusoidal oscillations of $\langle w(t) \rangle$ with steady amplitude following the initial decay. 
The period of oscillations of the current is estimated by $T_{w_0}/n$, where $T_{w_0}$ is the period of a superflow round trip in the absence of impurities~\cite{notaunits}.   
At a slightly higher impurity density ($n=8$), pinning dominates the early dynamics, causing $\langle w(t) \rangle$ to dip even below zero—signaling vortex emission beyond the initial winding number $w_0=6$, a mechanism that requires further investigation into the detailed mechanisms of vortex nucleation. 
After interacting with the impurities, the emitted vortices propagate at the outer edge. Notably, this configuration realizes a current switch: the initial clockwise motion ($w_0=6$) relaxes to a counter-clockwise circulation [$w(t\gg \tau) = -2$]. 
Finally, when the impurity density is sufficiently large [e.g., $n=10$ in Fig.~\ref{fig:Fig1}(b)], the vortex emission ceases, the winding number remains constant ($w(t)=w_0$), and the supercurrent is preserved over time, marking a stable configuration.
The enhanced stability effect for large $n$, shown in Fig.~\ref{fig:Fig1}(b), is a general phenomenon that we have observed numerically for different values of superfluid current.
The critical winding number $w_c$ separates stable from dissipative flow and increases with $n$, as shown in the phase diagram of Fig.~\ref{fig:Fig1}(c).
A similar stability enhancement was recently observed in toroidal traps with Josephson junction arrays~\cite{PezzeNATCOMM2024}. 
In our setup, impurities are smaller than the ring thickness, with the additional degree of freedom given by the impurity position, rather than the effective one-dimensional case of Josephson barriers~\cite{PezzeNATCOMM2024}.

%%%%%%%%%%%%%%%%%%%%%%%%%%%
%% Figure 3
%%%%%%%%%%%%%%%%%%%%%%%%%%%
\begin{figure*}[htb!]
\includegraphics[width=2.1\columnwidth]{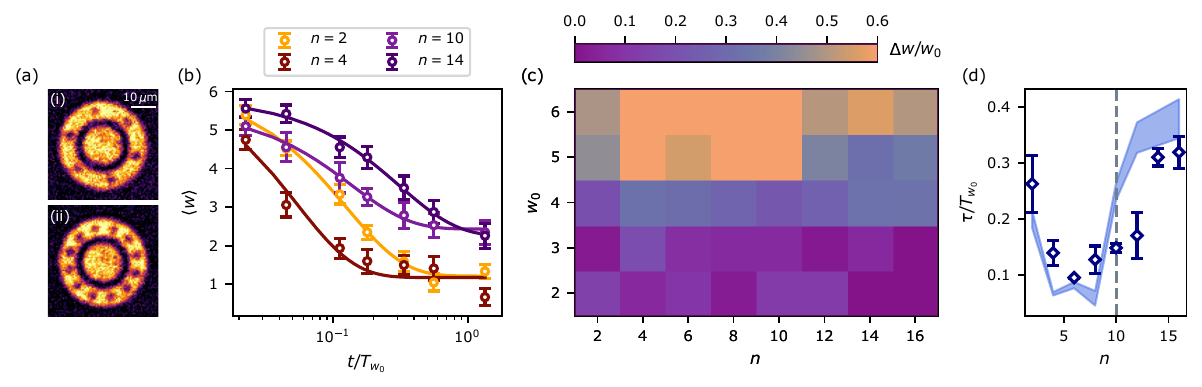}
\caption{{\bf Experimental results for impurities arranged at the ring center} 
(a) In situ images of the superfluid ring pierced by 6 (i) and 12 (ii) impurities of Gaussian shape with height $V_0/\mu = 3.7(7)$ and width $\sigma = 1.4(1) \, \mu$m
(see Appendix~\ref{AppB} for details). 
The central disk provides the phase reference for the interferometric probe of circulation. 
Each image corresponds to the average over $6$ independent realizations.
(b) Measured circulation as a function of time in rotation units \cite{notaunits} for $w_0=6$ and different number of impurities (see legend). 
Error bars here represent the standard deviation of the mean value over repeated realizations.
The solid lines are exponential fit of the experimental data, from which we extract the decay time $\tau$.
(c) Stability diagram for different initial circulation $w_0$ and number of impurities $n$. 
The color scale reports the relative circulation change $\Delta w/w_0 = (w_0-w_F)/w_F$, where $w_F$ is the average circulation measured at $t = 180 \,$ms.
(d) Decay time $\tau$ extracted from the fit as in panel (b), as a function of the number of impurities for $w_0=6$. 
Circles are experimental results, whereas the blue shaded area is the results of numerical simulations taking into account fluctuations in the impurities positions, widths and heights. 
The dashed gray line marks the transition value at which the system turns from unstable to stable for identical impurity, i.e., of same height, width and center's position. 
}
\label{fig:Fig3}
\end{figure*}
%%%%%%%%%%%%%%%%%%%%%%%%%%%
%%%%%%%%%%%%%%%%%%%%%%%%%%%
%%%%%%%%%%%%%%%%%%%%%%%%%%%

To understand the underlying enhanced-stability mechanism, we examine the local superfluid velocity $v(R^*,t)$, computed at the radius $R^*=\bar{r}-2\sigma$, between the impurity positions and near the ring’s inner boundary, where the velocity peaks and vortices first form [as shown in Fig.~\ref{fig:Fig1}(a)]. 
The velocity $v(R^*,t)$ reveals local maxima that oscillates slightly in time due to density fluctuations: we thus compute the time-averaged maximum velocity $\bar{v}_{\max}$ and compare it to the bulk speed of sound $\bar{c}_{\rm bulk} = \sqrt{g n_{\rm bulk} / m}$, where $g=4\pi\hbar^2 a/m$ is the interaction strength and $n_{\rm bulk}$ the bulk density, defined as the maximum density at radius $R^*$. 
The color scale in Fig.~\ref{fig:Fig1}(c) shows the time-averaged ratio $\bar{v}_{\max}/\bar{c}_{\rm bulk}$ for each $n$ and $w_0$ values.
For $w_0 \le w_c$, the time averages cover the full time evolution, while for $w_0 > w_c$ it accounts for only the times before the vortex emission. 
The critical velocity signaling the onset of dissipation closely matches the time-averaged $\bar{c}_{\rm bulk}$ and is nearly independent of impurity number.
Specifically, stable configurations satisfy  $\bar{v}_{\max}/\bar{c}_{\rm bulk} < 1$, near the critical point ($w_0=w_c$) $\bar{v}_{\max}/\bar{c}_{\rm bulk} \approx 1$ and exceed unit beyond it, signaling vortex emission and emergent dissipation. 
This confirms that dissipation typically begins when the flow velocity surpasses a critical value close to the bulk sound speed, consistent with an interpretation following the Landau criterion.
Figure 1(d) also shows that for fixed $w_0$, e.g., $w_0=5$, increasing impurity number $n$ reduces $\bar{v}_{\max}/\bar{c}_{\rm bulk}$, confirming that higher impurity density enhances stability. 

\textit{Asymmetric impurities configuration.}---We further investigate how a radial displacement $r_0$ of the impurities' position from the mid-radius $\bar{r}$ affects the stability and the vortex emission. 
Focusing on the case of six impurities, we vary $\Delta r/\xi = (r_0 - \bar{r}) / \xi$ from the ring center ($\Delta r = 0$) toward the inner ($\Delta r < 0$) and outer ($\Delta r > 0$) edges. 
We consider two scenarios [see Fig.~\ref{fig:Fig2}(a)]: shifting all impurities [(i) and (ii)] or moving just one of them [(iii) and (iv)]. 
Displacing all impurities toward the inner edge lowers the critical winding number $w_c$, while shifting them outward raises it, as illustrated by the green circles in Fig.~\ref{fig:Fig2}(b).
This trend follows a hydrodynamic reasoning \cite{DubessyPRA2012}: displacing the impurities outward reduces the maximum local velocity near the impurities [inset of Fig.~\ref{fig:Fig2}(b)], therefore higher $w_0$ is required to reach the critical flow speed $\sim \bar{c}_{\rm bulk}$.
Furthermore, when a single impurity is displaced, the innermost impurity sets the threshold, as shown by the gray dots in Fig.~\ref{fig:Fig2}(b).
In particular, $w_c$ decreases when moving the single impurity toward the center.
Instead, when moving the single impurity toward the outer edge of the torus, $w_c$ is determined by the $n-1$ impurities in the center and $w_c$ saturates. 
This configurational dependence also impacts vortex emission and the ensuing dynamics. 
While symmetric vortex emission is present in periodic impurity arrangements [Fig.~\ref{fig:Fig1}(a)], displacing a single impurity breaks this symmetry [Fig.~\ref{fig:Fig2}(a,iv)]. 
Furthermore, for $w_0>w_c$ when all impurities are moved radially, their position determines the vortex emission type: impurities near the center or inner edge emit single vortices, while those at the outer edge generate vortex dipoles—pairs of a vortex from the inner edge and an antivortex from the outer edge.
To quantify this dissipation, the evolution of the averaged winding number is modeled as an exponentially decaying function,
$\langle w(t) \rangle = w_0 e^{-t/\tau} + w_F$.
Impurities near the inner edge can quickly pin emitted vortices, which stay localized between the impurity and the zero-density center. 
This strong vortex–impurity interaction stabilizes the current: after a rapid initial drop, $\langle w(t) \rangle$ levels off at a higher final value $w_F$ [Fig.~\ref{fig:Fig2}(c-d)], indicating that only the segment of the ring between the inner edge and the impurity participates in phase slippage.
For $\Delta r/\xi = -4$ in Fig.~\ref{fig:Fig2}(d), vortices shuttle between the impurities and the central region, causing a slightly larger $\tau$ than for $\Delta r/\xi = -4$.
In contrast, impurities near the outer edge generate vortex dipoles [Fig.~\ref{fig:Fig2}(a–iv)], where antivortices can pin near the impurities during their dynamics, but vortices continue to propagate. 
This process contributes to a slower decay of $\langle w(t) \rangle$ reaching a lower final circulation $w_F$ [Fig.~\ref{fig:Fig2}(c-d)].

\textit{Experimental results.}---
Experimentally, we introduce a symmetric configuration of impurities at the center of the ring density by shaping a tailored repulsive optical potential using a digital micromirror device (DMD, see Appendix~\ref{AppB} for details). 
We monitor their effect on the initial current $w_0$ by interfering the ring with a central disk condensate [see Fig.~3(a)] \cite{DelPacePRX2022}. 
The experimental results confirm that the system's stability increases with the number of impurities, as illustrated in Fig.~3(b) for $w_0=6$.
Despite the high resolution of our experimental setup, allowing for positioning the impurities with a precision of a fraction of $\xi$ (see End Notes for details), the experimentally realized impurity configuration deviates from the perfectly symmetric case explored in Fig.~\ref{fig:Fig1}.
This is due to the spatial discretizations inherent to the DMD and the limitations on its optical projecting setup, hindering the experimental observation of a fully stable configuration.
However, the ratio $\Delta w_F / w_0 = (w_0-w_F)/w_0$, where $w_F$ is the measured circulation at $180\,$ms, reveals a clear increase of stability for increasing number of impurities for large $w_0$, Fig.~\ref{fig:Fig3}(c), in qualitative agreement with the theoretical analysis of Fig.~\ref{fig:Fig1}(c).
The same behavior is also manifest in the trend of the fitted decay time $\tau$ reported in Fig.~\ref{fig:Fig3}(d) for $w_0=6$, which,
after a first drop for small $n$, increases again for large impurity numbers. 
This behavior is confirmed by numerical simulations of the current stability in the presence of impurities with properties randomly chosen in the range of the experimental uncertainty (see Appendix~\ref{AppB}).
Also the numerical simulations predict an increased stability for large $n$,  without ever recovering a totally stable supercurrent as for the ideal case of identical and symmetric distribution of impurities of Fig.~\ref{fig:Fig1}.
Deviations from theoretical behavior may be due to the finite temperature of the experiment, which nevertheless does not preclude the observation of the increased stability.

\textit{Conclusions and outlook.---}This work focused on the stabilization mechanisms induced by multiple impurities in ring superfluids.
Future studies should systematically investigate and probe the opposite, unstable, regime, where the interplay of vortex pinning and unpinning could be exploited to program the superflow in the ring. 
Potential applications include atomtronic DC-to-AC inverters, exploiting the vortex dynamics to induce oscillations of the superfluid current, and atomtronic current reversers, that use the simultaneous emission of $n$ vortices to convert clockwise into counterclockwise flow, as shown in Fig.~\ref{fig:Fig1}(b). 
At present, these engineering possibilities are hindered by the small (sub-micron) uncertainties in impurity positions, as our analysis shows. 
Future experimental and theoretical studies could address whether these limitations can be overcome by employing larger impurities or engineered weak links in tailored arrangements, and by accounting for finite-temperature effects, which are expected to modify vortex emission and pinning.\\

\begin{acknowledgments}
\textit{Acknowledgments.---} We thank Brynmor Haskell for fruitful discussions and Cyprien Daix for contributing to the initial experimental activity in the early stage of this work. 
K. X. acknowledge funding from the Italian MUR (PRIN DiQut Grant No. 2022523NA7).
G.R. and G.D.P. acknowledge financial support from the PNRR MUR project PE0000023-NQSTI. 
G.R. acknowledges funding from the Italian Ministry of University and Research under the PRIN2017 project CEnTraL and project CNR-FOE-LENS-2024. 
The authors acknowledge support from the European Union - NextGenerationEU within the “Integrated Infrastructure Initiative in Photonics and Quantum Sciences" (I-PHOQS). 
The authors acknowledge funding from INFN through the RELAQS project. 
This publication has received funding under the Horizon Europe programme HORIZON-CL4-2022-QUANTUM-02-SGA (project PASQuanS2.1, GA no.~101113690) and Horizon 2020 research and innovation programme (GA no.~871124).
\end{acknowledgments}

\section{Appendix}

\subsection{Numerical methods}
\label{AppA}

We have simulated the dynamics of a BEC of $^6$Li molecules in a toroidal trap by solving numerically   the mean-field Gross-Pitaevskii equation (GPE):
\begin{equation}
    i\hbar \frac{\partial  \psi (\mathbf{r},t)}{ \partial t}=\frac{-\hbar ^2}{2M} \nabla ^2 \psi (\mathbf{r},t)+V(\vect{r}) \psi(\mathbf{r},t) +g |\psi(\mathbf{r},t)| ^2 \psi(\mathbf{r},t)
    \label{eq.GP.methods}
\end{equation}
with $\psi(\mathbf{r},t)$ being the condensate wave function,  $M=2m_{^6{Li}}$ the lithium molecule mass,  $V$ the external trapping potential, $g=4 \pi \hbar^2 a/M$ the interaction strength between molecules. 
The trapping potential is given by the sum of a harmonic confinement
$V_{\mathrm{harm}}= m ( \omega_\perp^2 r^2 + \omega_z^2 z^2)/2$, with $\{\omega_\perp , \, \omega_z \}= 2\pi \, \times \{2.5 \, , \, 396\}$ Hz the radial and axial trapping frequencies, respectively, and a ring trap 
\begin{equation}
\displaystyle
V_{\rm ring}=V_1 \left[\tanh\left(\frac{r-R_{\mathrm out}}{\sigma _1}\right)+1\right]+V_1 \left[\tanh\left(\frac{R_{\rm in}-r}{\sigma _1}\right)+1\right].
\label{Vbound}
\end{equation}
Here, $R_{\mathrm in}=9.6\mu \text{m}$ and $R_{\mathrm out}=21 \mu \text{m}$ 
are the inner and outer radius, respectively, and 
the parameter $\sigma _1=0.37 \mu \text{m}$ characterizes the stiffness of the hard walls
with $V_t > \mu$ with $\mu$ the chemical potential so the density goes to zero at the boundary. The molecule number is $N=5100$ leading to a chemical potential value $\mu=860$ Hz and the healing length $\xi=\hbar/\sqrt{2\mu M} = 0.68 \mu$m. Additionally the impurities potential is summed to the external potential with each impurity being modeled by a Gaussian potential periodically distributed along the ring: 
\begin{align}\label{eq:potential}
V_{\textrm{d}}(\bf{r}) &=  V_0\sum_{i=1}^{n}\exp\!\left[-\frac{(x-r_0\cos(2\pi i/n))^2+(y- r_0\sin(2\pi i/n))^2}{2\sigma^2}\right],
\end{align}
with $V_0$ the impurity height and $\sigma$ its $1/e^2$ width. 

Equation~\ref{eq.GP.methods} is solved numerically by the Fourier split-step method on a Cartesian grid of $\{ N_x , N_y , N_z \} = \{ 256 , 256 , 80\}$ points dividing  a grid size of length $-34.846 \, \mu \text{m} \le r \le 34.846 \,  \mu \text{m}$ and $-11.0 \, \mu \text{m}  \le z \le 11.0 \, \mu \text{m}$ in the radial plane and axial direction, respectively. 
The time step is instead set to $\Delta t = 1 \times 10^{-5}\, \omega_\perp^{-1}$. 

We initially find the system ground state by solving the GPE in imaginary time in the presence of the $n$ impurities, with the wavefunction being multiplyied with the phase factor exp($-i 2 \pi w \theta$) where $\theta$ is the azimuthal angle. In this way we imprint a supercurrent with initial winding number $w_0$, which is persistent  in the absence of impurities. 
Then we study the current dynamics by solving the time-dependent GPE and by extracting the winding number and superfluid velocity temporal evolution. We implement three main impurities configuration in our simulations: (i) symmetric impurities configuration with impurity height $V_0=3.7 \mu$, width  $\sigma$ =1.4$\mu$m$\simeq 2\xi$ and distributed along the mid-ring radius $\bar{r}$ (ii) same impurities parameters as (i) but with their center different from $\bar{r}$ and (iii) including the experimental fluctuations in the numerical simulations by randomly extracting the impurity height, width and position from three Gaussian distributions with same mean and standard deviation as the experimental impurity properties. The results shown in Fig.~3 are the average of five numerical realization for each impurity number $n$.

\subsection{Experimental methods}
\label{AppB}

The experimental study of the current stability in a molecular Bose-Einstein condensate of Lithium-6 atoms is performed following a similar procedure as the one described in Ref. \cite{PezzeNATCOMM2024}. Specifically, we realize a ring condensate of $N \simeq 6 \times 10^3$ atom pairs, with approximately 85\% condensed fraction and a molecular scattering length of $a_M \simeq 1030 \, a_0$. The BEC is confined along the vertical by a tight harmonic trap of trap frequency  $\nu_z = 396\,$Hz.  %Under these parameters, the condensate chemical potential is $\mu = 860 \,$Hz and the healing length $\xi = 0.68 \, \mu$m.

We inject the desired current at winding $w_0$ in the ring by phase imprinting and then ramp up in $2.6$ ms the impurity repulsive optical potential. This is created by sculpting a 532 nm laser intensity profile with a Digital Micromirror Devide (DMD) and imaging it on the atomic plane via a high-resolution ($\sim 1 \, \mu$m) optical setup. The same setup also produced the trap geometry combining the ring with the central disk and the azimuthal gradient profile used for the phase imprinting. 
We probe the current by interfering the ring superfluid with the central disk, acting as a phase reference. In particular, to monitor the current dynamics in the presence of the impurities, we first ramp off the impurity optical potential in $2.6\,$ms and then proceed with the interferometric detection of the current.
For a given number of obstacles in the ring, we measure the average circulation $\langle w(t)\rangle$ for various interaction times, calculating the average over 15 images. We then perform a fit of the time evolution using the function $w(t) = w_f+A\exp (-t/\tau)$, as shown in figure \ref{fig:Fig2}. This procedure allows us to extract a decay time for each obstacle configuration.

We characterize the properties of the impurities by acquiring an image of the DMD-made light intensity profile using a service CCD camera. The image of each impurity is fitted with a 2D Gaussian function to extract its depth $V_0$, width $\sigma$ and radial position $r_0$. These properties are studied for all impurity configurations $n\leq 16$, providing the average quantities: $V_0/\mu = 3.7(7)$, $\sigma = 1.4(1) \, \mu$m and $r_0 = 14.9(2) \, \mu$m. Note that the size $\sigma$ reported here corresponds to the average size of the 2D Gaussian along the two directions, as the average impurity aspect ratio is $1.0(1)$. The numerical simulations reported in Fig.~3 have been performed by randomly extracting each impurity properties from the confidence range of the experimental parameters reported above. Remarkably, despite the relative larger uncertainty in the impurity height ($20\%$), it is the sub-micron positioning uncertainty $<2\%$ to mostly affect the current stability. In fact, the impurities considered in this study have a height well above the superfluid chemical potential, and its variations are irrelevant for the supercurrent stability as long as the condition $V_0/\mu >1$ is satisfied.


\begin{thebibliography}{100}
\bibitem{notaunits}
We indicate with $T_{w_0}$ the time the superfluid takes to make a loop around the ring at mid-radius $\bar{r}$ and in the absence of impurities.
That is: $T_{w_0} = 2 \pi \bar{r}/v(\bar{r}) = 2 \pi \bar{r}^2 m/(\hbar w_0)$, where $v(\bar{r}) = \hbar w_0/(m \bar{r})$ is the superfluid speed at $\bar{r}$ for an initial circulation $w_0$.
%
For the parameters of Fig.~\ref{fig:Fig1}(a) and (b), namely $w_0=6$ and $\bar{r}=15.3$ $\mu$m, we have $T_{w_0} \approx 46$ ms. In Fig.~\ref{fig:Fig2}(d), for each point we account for the corresponding $T_{w_0}$ set by the initial winding $w_0$.
\end{thebibliography}
\end{document}